\documentclass[a4paper,11pt]{article}
\pdfoutput=1

\usepackage{jinstpub}
\usepackage{amsmath,bm}
\usepackage{amsfonts}
\usepackage{amssymb}
\usepackage{graphicx}
\usepackage{multirow}
\usepackage{units}
\usepackage{siunitx}
\usepackage{url}
\usepackage{subcaption}
\usepackage{lineno}

\title{First observations of speed of light tracks by a fluorescence detector looking down on the atmosphere}

\author{The JEM-EUSO Collaboration}

\emailAdd{jeser@mines.edu}

\abstract{EUSO-Balloon is a pathfinder mission for the Extreme Universe Space Observatory onboard the Japanese Experiment Module (JEM-EUSO). It was launched on the moonless night of the 25$^{th}$ of August 2014 from Timmins, Canada. The flight ended successfully after maintaining the target altitude of \SI{38}{\kilo\meter} for five hours. One part of the mission was a 2.5 hour underflight using a helicopter equipped with three UV light sources (LED, xenon flasher and laser) to perform an inflight calibration and examine the detectors capability to measure tracks moving at the speed of light. We describe the helicopter laser system and details of the underflight as well as how the laser tracks were recorded and found in the data. These are the first recorded laser tracks measured from a fluorescence detector looking down on the atmosphere. Finally, we present a first reconstruction of the direction of the laser tracks relative to the detector.}


\begin{document}
\maketitle
\flushbottom

\section{Introduction}
\label{sec:intro}
To measure extreme energy cosmic rays with high statistics a large observation area is needed. One option is to go to space. JEM-EUSO (Extreme Universe Space Observatory on board the Japanese Experiment Module) is a planned fluorescence detector on the International Space Station \cite{JEMEUSO}. It is designed to measure the light of extensive air showers developing in the Earth's atmosphere beneath the detector. Further space instruments in the design stage are KLYPVE-EUSO \cite{KEUSO} and POEMMA \cite{POEMMA}. Various prototypes are being developed for the JEM-EUSO mission \cite{EUSOTA,ICRC2017:Lawrence,MiniEUSO}. The first one looking down onto the atmosphere, was EUSO-Balloon.\\ 
The EUSO-balloon mission had three main objectives: 
\begin{enumerate} 
\item Perform an end-to-end test of the JEM-EUSO design in a near-space environment 
\item Measure the effective terrestrial UV background relevant for all space-based fluorescence detectors (a discussion can be found in \cite{ICRC2015:SimonKenji}) 
\item Detect UV light from above including laser tracks for the first time.
\end{enumerate} 
The third one is an important milestone for space-based fluorescence measurements. The detector was flown as a stratospheric balloon payload during the moonless night of the 25$^{th}$ of August 2014. It was launched from the Timmins (Canada) stratospheric balloon launch facility. An essential part of the mission was a 2.5 hour underflight using a helicopter equipped with three UV light sources. The light sources were used to perform an inflight calibration and to test the instrument's detection capabilities. The EUSO-Balloon instrument and mission have been reported elsewhere \cite{ICRC2015:Peter}. We will focus on the laser part of the underflight. First a brief description of the system and the idea behind the underflight is presented. Then we show an example of the obtained data. Finally, we will explain how the direction of these tracks can be reconstructed and discuss the results and impact of this reconstruction.\\
\begin{figure}
   \begin{subfigure}{.5\columnwidth}
     \centering
     \includegraphics[width=1\linewidth]{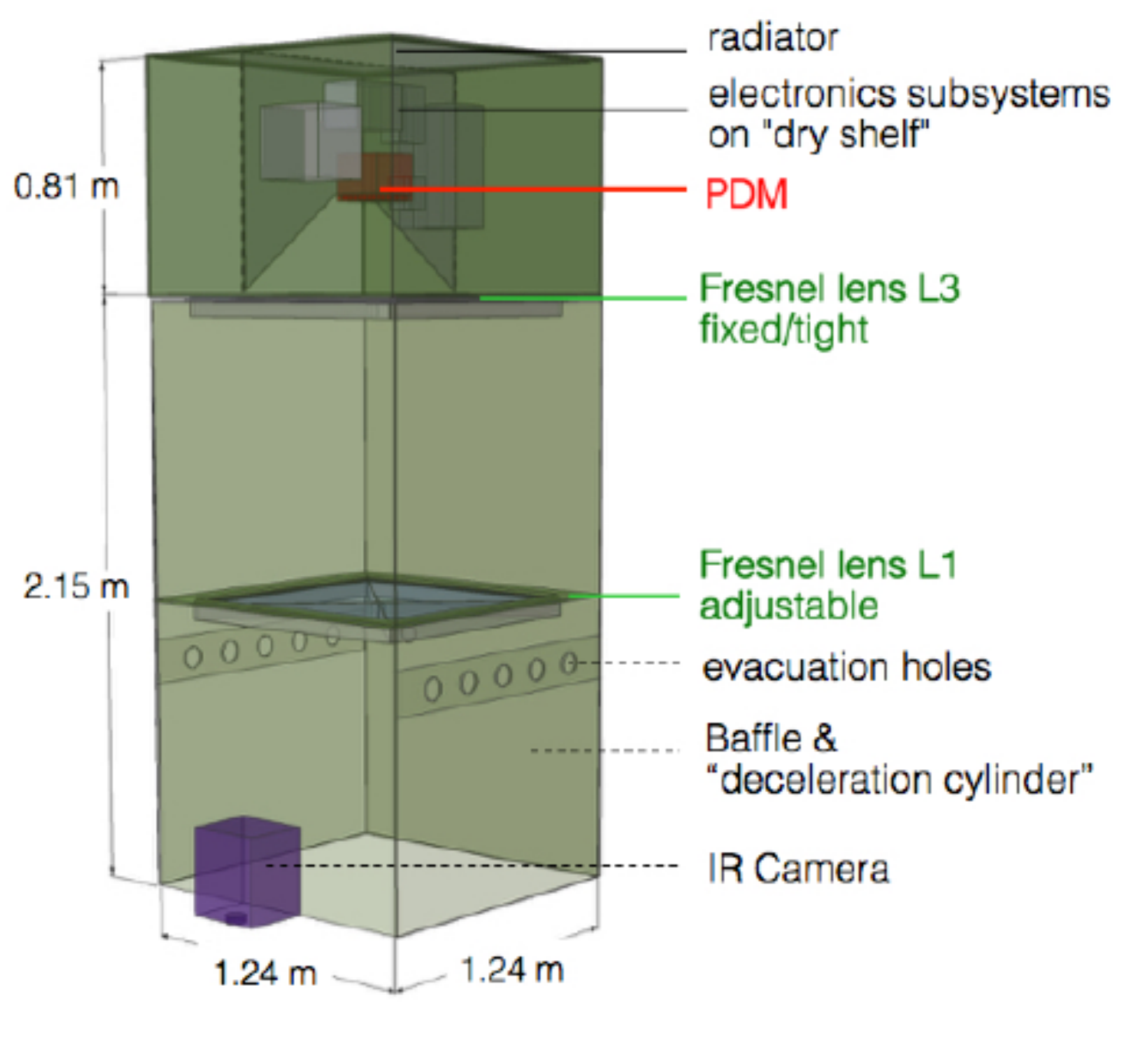}
   \end{subfigure}%
   \begin{subfigure}{.5\columnwidth}
      \centering
      \includegraphics[width=.6\linewidth]{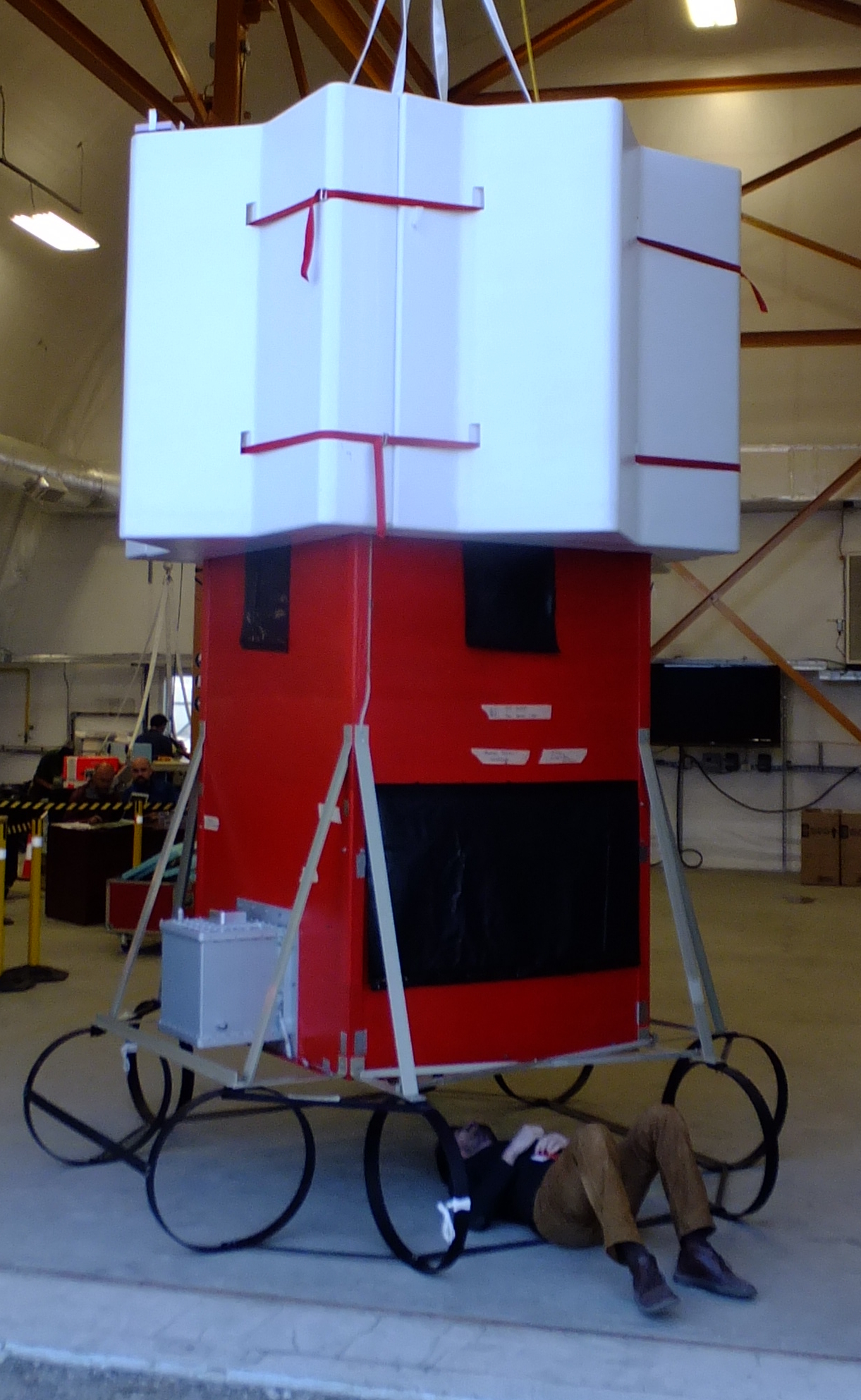}
   \end{subfigure}   
   \caption{left: schematic of the flown EUSO-Balloon detector \cite{ICRC2015:Peter}; right: the actual detector.}
   \label{fig:payload}
  \end{figure}
\section{Detector}
\begin{figure}
     \centering
     \includegraphics[width=.5\linewidth]{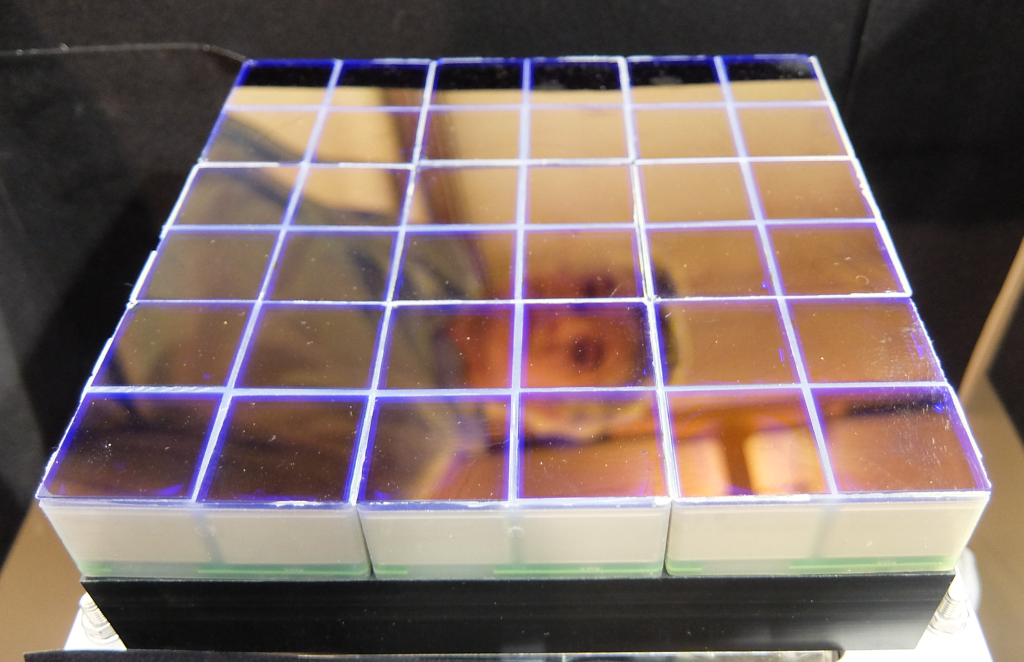}
     \caption{Top view picture of the PDM with filters attached}
     \label{fig:PDM}
\end{figure}

An overview of the detector is shown in fig. \ref{fig:payload} and table \ref{tab:DetProperties} lists its key properties. The instrument is a high speed UV camera designed to measure the fluorescence light of cosmic ray air showers. The two main components are the optical bench and the instrument booth. The optical bench contains two Fresnel lenses with an aperture of \unit[1]{m$^2$} to focus the light arriving at the instrument's aperture onto the Photo Detection Module (PDM) of EUSO-Balloon. The point spread function (PSF) of the optics was defined as the FWHM of a two-dimensional Gaussian fit to the focal point. For EUSO-Balloon this gives \SI[separate-uncertainty = true,multi-part-units=single]{9.0(2)}{\mm} or 3$\times$3 pixels corresponding to \SI{0.7}{\degree} $\times$ \SI{0.7}{\degree}. EUSO-Balloon had a field of view of \unit[$\pm$5.5]{$^\circ$}. A detailed description of the performance of the optical system is given in \cite{ICRC2015:Catalano}.\\
The PDM (see fig. \ref{fig:PDM}) is made of 36 Hamamatsu M64 Multi-Anode Photomultiplier Tubes (MAPMT), each containing 64 anodes (2304 pixels in total) capable of single photoelectron counting \cite{PDM}. The Schott BG3 optical filter leads to a detection band of 290 to \unit[430]{nm} with a tail up to \SI{500}{\nano\meter}. The time binning of the detector is \SI{2.5}{\mu\s} (equivalent to 1 Gate Time Unit, GTU). One event trigger causes 128 GTUs of data to be collected from all pixels. The trigger rate during the flight was \SI{20}{\Hz}, set by an internal clock. There was no synchronization between this trigger and the laser system.\\
\begin{table}[htp]
\centering
\caption{Specifications of EUSO-Balloon and 2014~mission.}
\label{tab:DetProperties}
\begin{tabular}{| l| l| l l|}
\hline
& Specification & Notes&\\
\hline
Telescope Optics & 2$\times$ 1 m$^2$ Fresnel lenses& PMMA &\\
Field of View & 11${^\circ}$${\times}$11${^\circ}$ & &\\
Number of Pixels & 2304 (48${\times}$48)& 36~ 64 ch. MAPMTs &\\
MAPMT & R11265-113-M64-MOD2~& Hamamatsu &\\
UV Filter & BG3,~2~mm thick& 1 per MAPMT &\\
Read Out & DC coupled & double pulse separation 30~ns \cite{ASIC} &\\
Time Bin Duration & 2.5~${\mu}$s (GTU) & event packet = 128~bins (320~${\mu}$s) &\\
& 2.3~${\mu}$s integration + 0.2~${\mu}$s dead time & &\\
Trigger & forced by CPU & \unit[20]{Hz}, non-synchronized &\\
Flight CPU & Atom N270 1.6 GHz processor~ & Intel &\\
Telemetry & ${\approx}$ 1.3~Mbits/s & NOSYCA CNES& \\
Power Consumption & 70~W & &\\
Detector Weight & 467 kg & & \\
\hline
Balloon &4.0${\times}10^{5}$ m$^{3}$ & helium & \\
Nominal Float Height & 38300 m & &\\
Launch &August 25 00:53~UTC 2017~& \SI{48.57}{\degree}N~lat \ \ \SI{81.38}{\degree}W~long&\\
Flight Duration & 8 hours & & \\
\hline
\end{tabular}
\end{table}

\section{Laser system and motivation}
\label{sec:LaserSys}
\begin{figure}
   \center
   \includegraphics[width=.65\columnwidth]{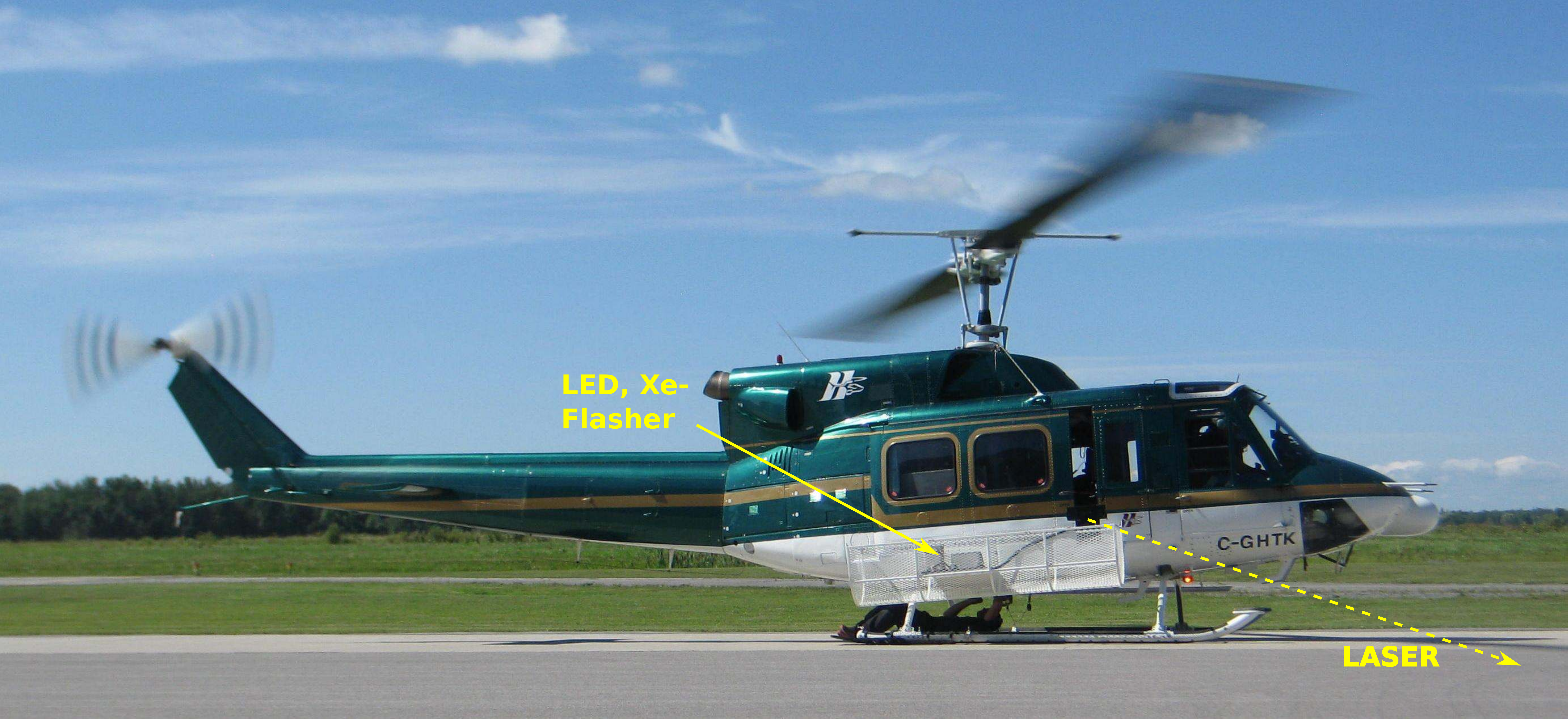}
   \caption{The light sources were mounted on a Bell 212 helicopter.}
   \label{fig:Helicopter}
\end{figure}

\begin{figure}
   \center
   \includegraphics[width=.9\columnwidth]{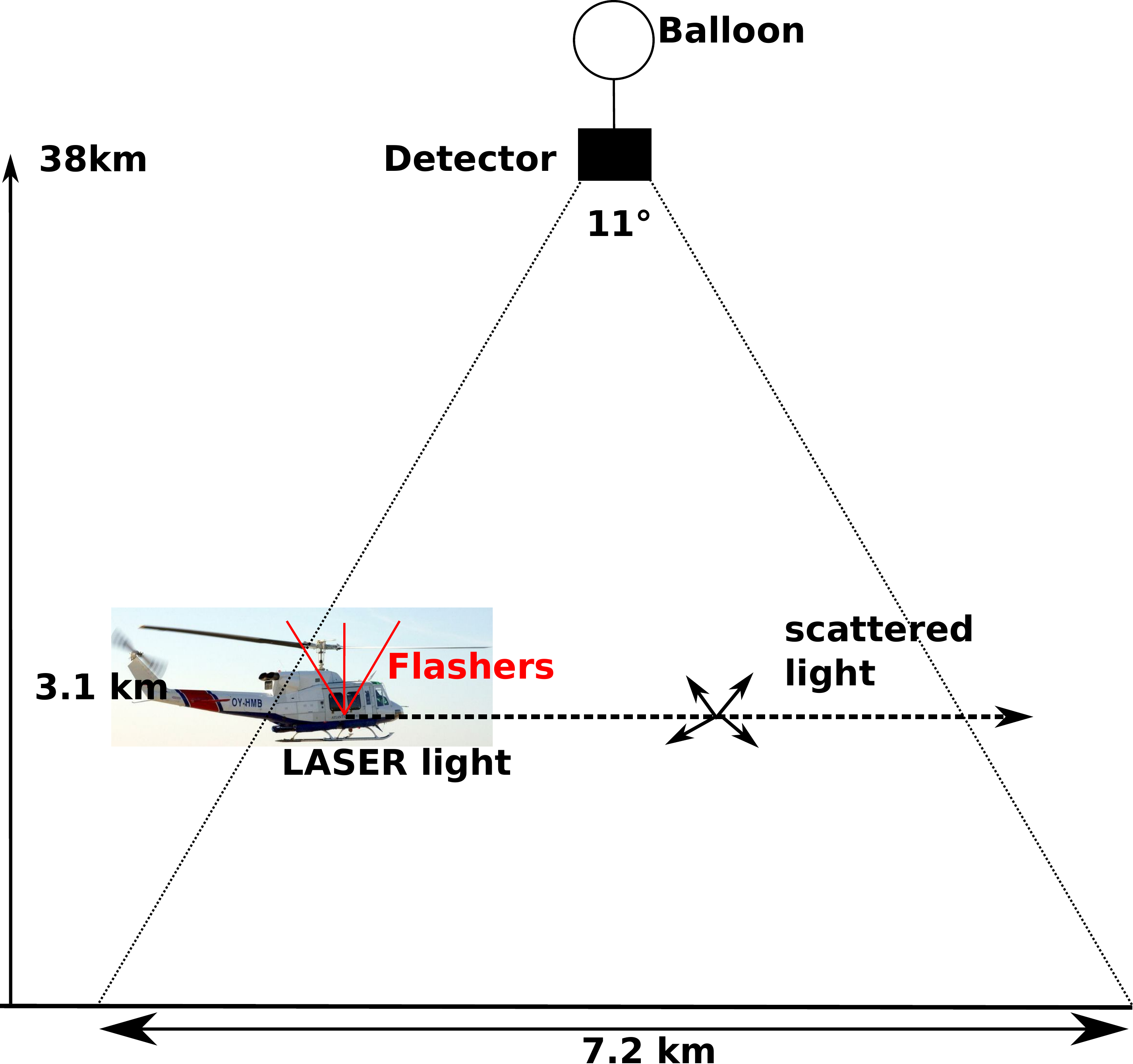}
   \caption{Sketch of the helicopter underflight}
   \label{fig:basicIdea}
\end{figure}

\begin{figure}[h]
  \center
   \begin{subfigure}[b]{0.5\textwidth}
   \centering
   \includegraphics[width=\linewidth]{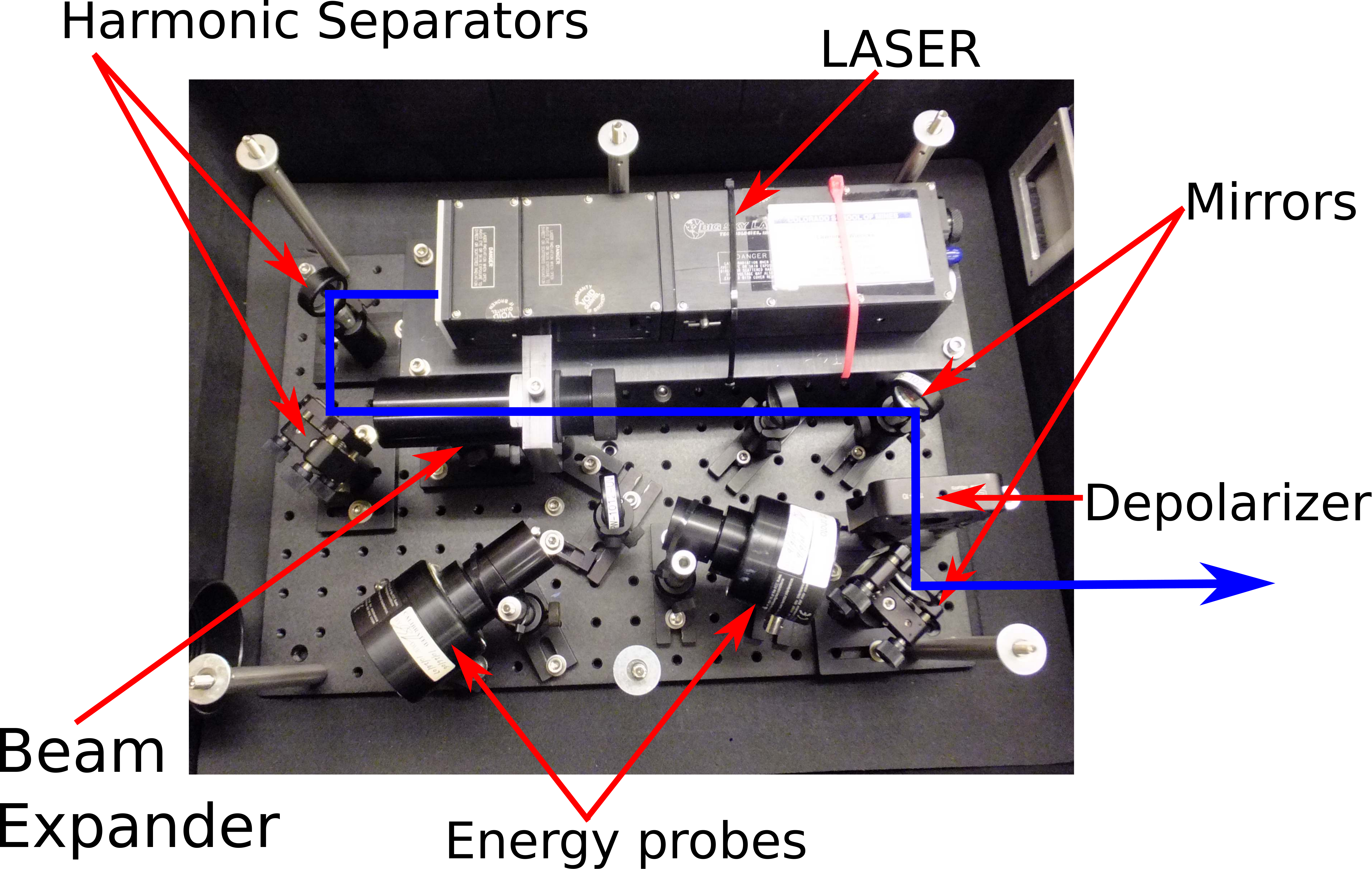}
   \caption*{}
   \end{subfigure}
   \hfill
   \begin{subfigure}[b]{0.4\textwidth}
   \centering
   \includegraphics[width=\linewidth]{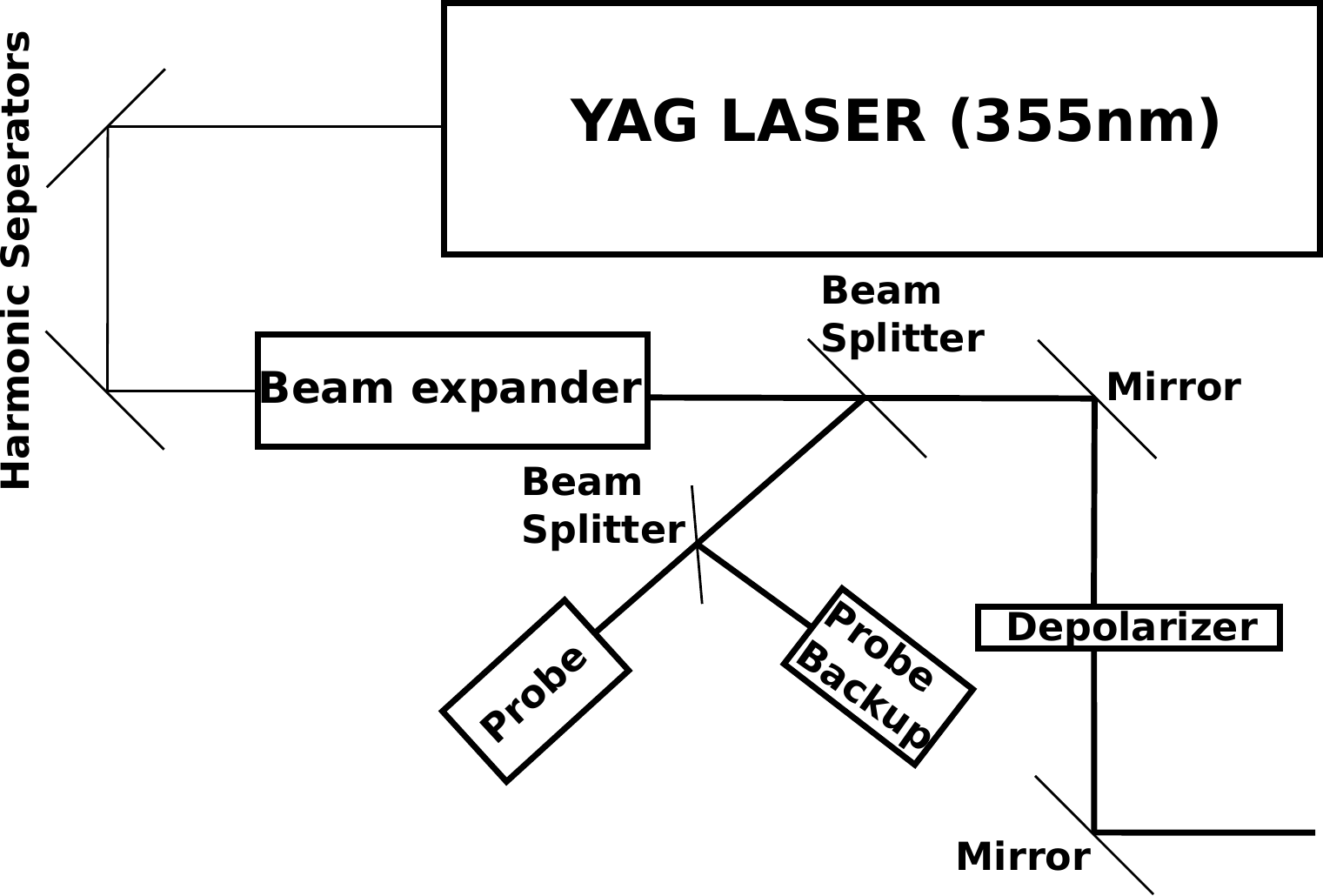}
   \caption*{}
   \label{fig:LaserBoxDrawing}
   \end{subfigure}
   \caption{Left: The UV-laser system that was flown in the helicopter. The thick blue line 
   indicates the beam path. Right: Schematic of the UV-laser system}
   \label{fig:laserSys}
\end{figure}

To evaluate the detectors capability to measure light from an Extensive Air Shower (EAS) and to perform an in-flight calibration, three light sources were mounted to a helicopter that flew under the balloon (see fig. \ref{fig:Helicopter}). An illustration of the underflight arrangement is shown in figure \ref{fig:basicIdea}. The three light sources were: a UV-LED, a xenon flashlamp, and a UV-laser.
Light is scattered isotropically out of a randomly polarized, pulsed laser beam when shot into the atmosphere. This light can be recorded by a fluorescence telescope used to look for cosmic rays. Both air shower and laser, produce a track moving with the speed of light that is observed by the detector. Unlike showers it is possible to set the laser energy and direction and repeat the measurement whenever needed. This makes a laser a perfect test beam for fluorescence cosmic ray detectors. A more comprehensive study is presented in \cite{CalosThesis}.
\begin{figure} 
     \centering   
     \includegraphics[width = .75\textwidth]{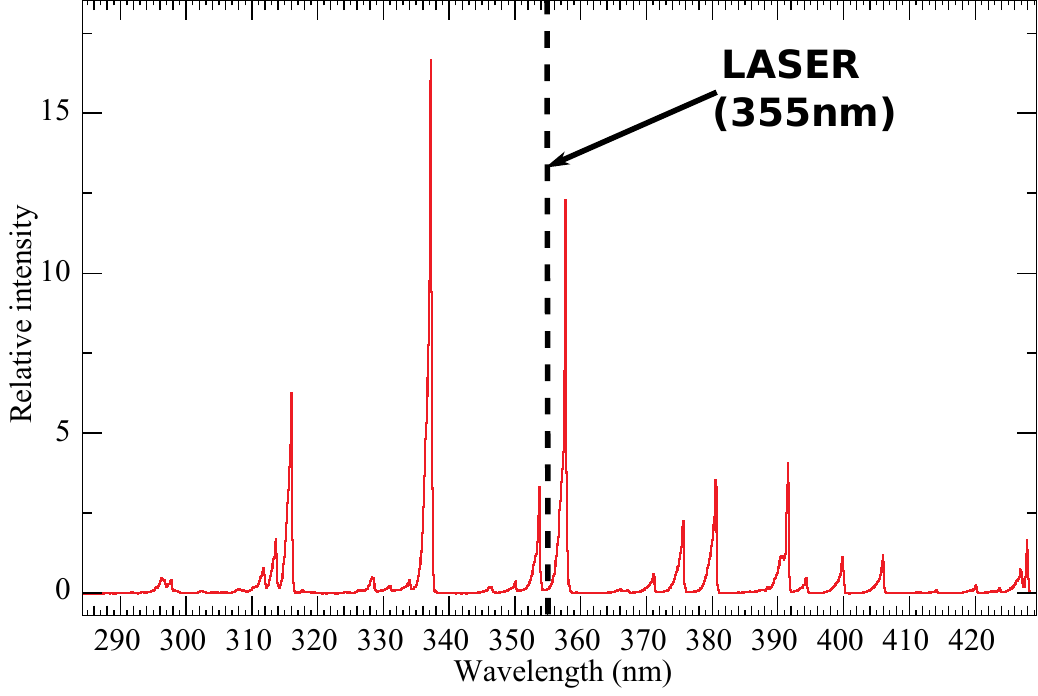}%
     \caption{The wavelength of the laser (355 nm) is indicated on the fluorescence spectrum of electrons in air \cite{Ave:2007xh}.}
     \label{fig:FluoSpectrum}
\end{figure}
This paper will focus on the laser as a light source. Additional details of the LED and the Flashlamp can be found in \cite{ICRC2015:Jim}.
The laser used was a Quantel CFR-Ultra \cite{WWW:Quantel} YAG-laser with frequency tripling to a wavelength of \SI{355}{\nm}. This wavelength was chosen because it is in the middle of the atmospheric fluorescence spectrum (see fig. \ref{fig:FluoSpectrum}). Its maximum energy is \SI{18}{mJ} with a \SI{7}{ns} pulse width. The optical setup is shown in fig. \ref{fig:laserSys}. Harmonic separators are used to achieve a spectral purity of more than 99.9\%. A 3$\times$ beam expander reduces the divergence to less than \SI{0.04}{\degree}. The beam splitter diverts 5\% of the primary beam onto a pyroelectric probe \cite{WWW:LaserProbe} that measures the relative energy for every discharge. The depolarizing optics is used to randomly polarize the laser beam. This way the scattering out of the beam in air would be symmetric in the azimuth angle around the beam axis.\\
The beam characteristics are listed in table \ref{tab:BeamChar}. The laser system was calibrated before and after the flight by measuring the ratio between the monitor energy probe and a second energy probe placed temporarily in the beam downstream of all optics. The calibration factor is independent from the stability of the laser itself. The difference between this calibration factor measured before and after the flight was 1.2 \%.
\begin{table}[h]
   \caption{Laser beam characteristics}
   \label{tab:BeamChar}
   \center
   \begin{tabular}{|c|c|c|c|}
   \hline
    wavelength & \unit[355]{nm} & divergence & $<$ \unit[0.04]{$^{\circ}$}\\ \hline
    relative energy calibration & $<$ 2\% & beam halo & $<$ 0.5\%\\ \hline
    absolute energy calibration & $<$ 4.5\% & spectral purity & $>$ 99.9\% \\ \hline
    overall stability & 1.2\% & \multirow{2}{*}{beam pointing direction} & zenith angle: $ < \unit[3]{^{\circ}}$\\ 
    \cline{1-2}
    depolarization & $<$ 4\% & & azimuth angle: $ < \unit[1]{^{\circ}}$ \\ \hline
     absolute timing accuracy  & $\pm$\unit[20]{ns} & repetition rate (GPS sync.) & $\unit{19}{Hz}$\\ \hline
   \end{tabular}
\end{table}
The laser system is controlled by a Single Board Computer (SBC). The key component of the SBC is the "GPSY2" module \cite{GPSY}. It contains an on-board GPS receiver, two \unit[5]{V} outputs, and one analogue input which triggers the readout of all components. The timing accuracy is \SI{100}{\ns}. The SBC was used to trigger the light sources in a specific order (1. LED, 2. Laser, 3. Xe-Flasher). The sequence was timed so that light from the three sources could reach the detector in the same 128 GTU readout window.\\ 
The system was mounted within a Bell 212 helicopter. The laser beam was fired through a partially open door, perpendicular to the body of the helicopter and horizontally when the helicopter was flying level. The pointing accuracy of the laser is better than \SI{1}{\degree} in azimuth and better than \SI{3}{\degree} for the zenith angle. The pointing direction was assessed using a self leveling laser while the system was mounted in the helicopter.

\section{Flight summary}
\label{sec:FlightSummary}
\begin{figure}[h]      
   \centering
   \includegraphics[width=.75\columnwidth]{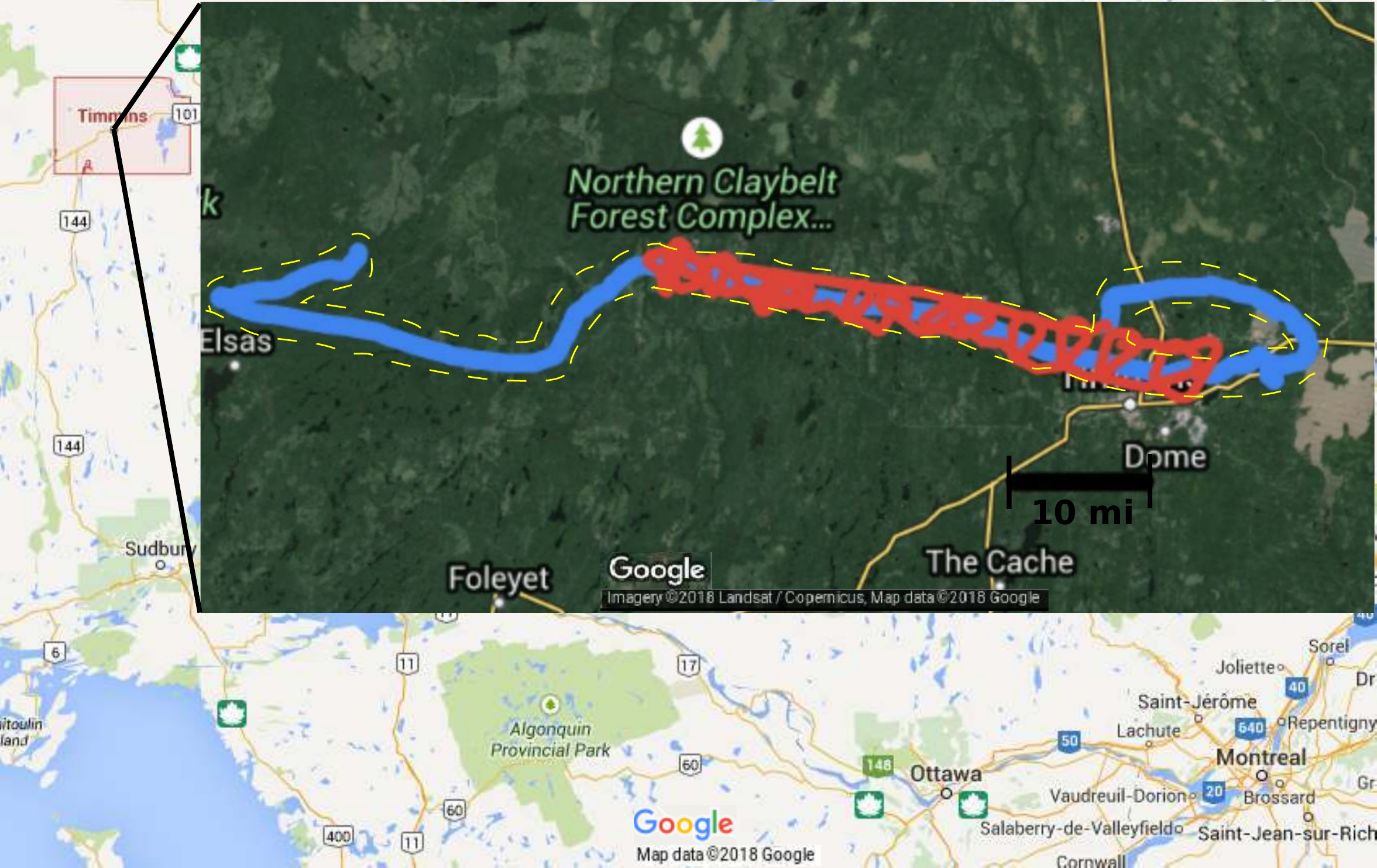}
   \caption{The thick blue line shows the flight path of the balloon. The yellow dashed line is the 
   approximate 
   FoV. The red line (loops) shows the flight pattern of the helicopter while the light sources 
   were firing (\textit{created with Google maps, Imagery Landsat/Compernicus, Map data Google)}.}
   \label{fig:FlightPath}
\end{figure}

\begin{figure}   
   \centering
   \includegraphics[width= .75\textwidth]{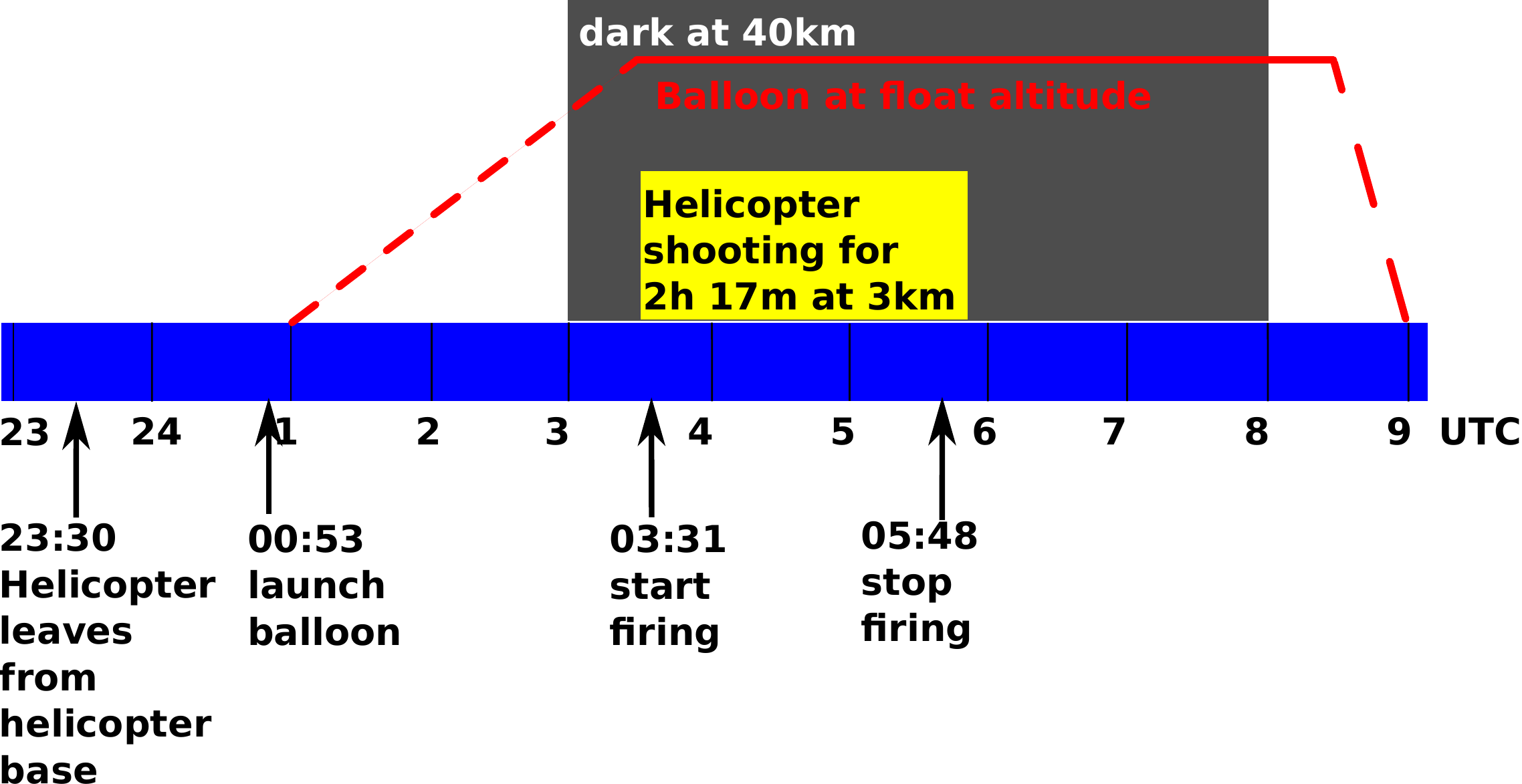}
   \caption{Timeline of the balloon and helicopter mission. The red line indicates the balloon.}
   \label{fig:TimeLine}  
\end{figure}

The balloon was launched on the 25$^{th}$ of August 2014 at 00:53 UTC and reached a float altitude of approximately 38 km after an ascent of 2h 50m. After 4h 40m at float altitude the payload was released from the balloon (8:20 UTC) and landed 39 minutes later in a lake. The trajectories of the balloon and the helicopter are shown in fig. \ref{fig:FlightPath} and the timeline of the mission is shown in figure \ref{fig:TimeLine}. The flight path of the helicopter is plotted in red in figure \ref{fig:FlightPath}.\\
For the underflight the helicopter arrived from Ottawa (helicopter base, \SI{550}{\km} from Timmins) on the evening of the launch. After refueling in Timmins, it followed the balloon using a GPS tracking system. The tracking system consisted of a GPS module and a beacon mounted on the balloon and a receiver on board of the helicopter \cite{ICRC2015:Jim}. At 03:31 UTC the helicopter entered the FoV of the detector and the light sources were turned on. At the height of the helicopter of \SI{3}{\kilo\meter} the FoV of the detector is 6.7 by 6.7 kilometers. To trigger pixels across the whole PDM, the pilots flew circular loops with a radius smaller than \SI{4}{\km} centered on the position of the balloon while the laser was pointing towards the center of the circles. The sources were fired $\sim$ 150000 times in 2.28 hours.\\

\section{Data collection and example events}
\label{sec:ExampleEvent}
\begin{figure}[h]
   \center
   \includegraphics[width=.8\columnwidth]{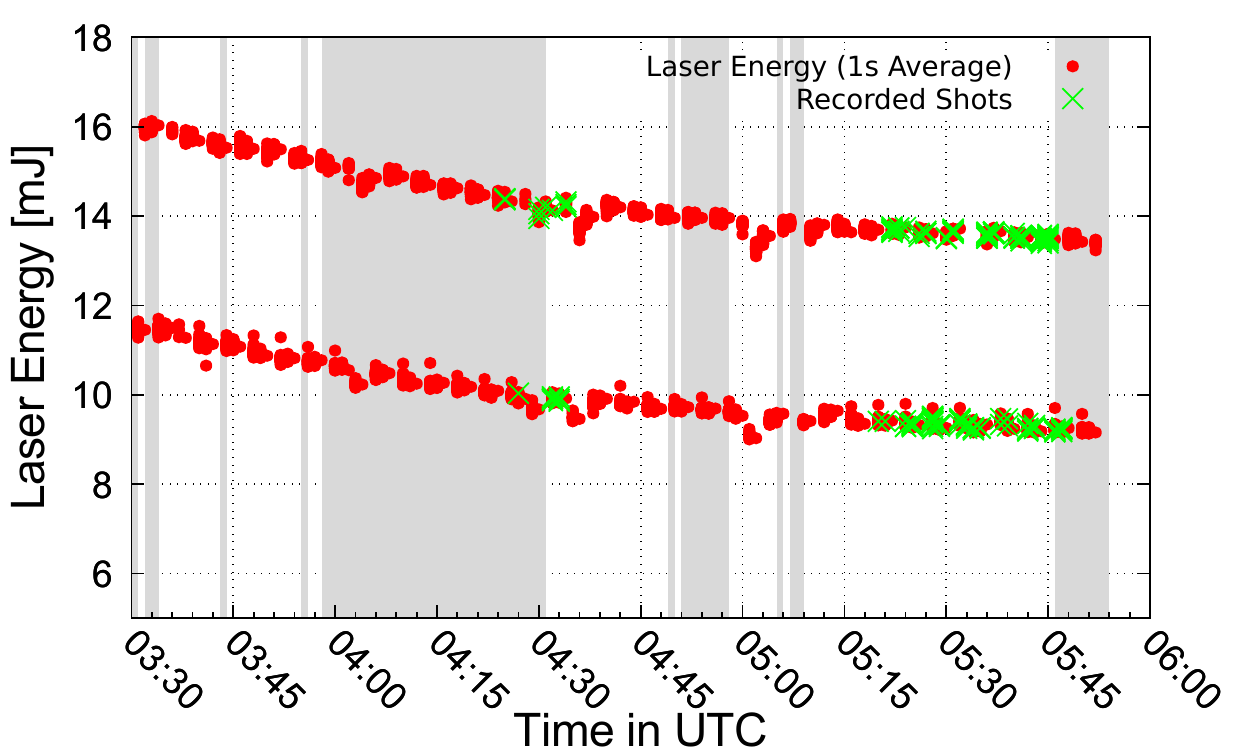}
   \caption{Red dots: Energy of all fired laser shots averaged over 19 shots (\SI{1}{\second}). Green Xs: Shots 
   recorded by EUSO-balloon. Grey regions indicate the likely presence of clouds. The laser energy is decreasing due to heating of the laser itself.}
   \label{fig:LaserShots}
\end{figure}

\begin{figure}[h]
   \center
   \includegraphics[width=.75\columnwidth]{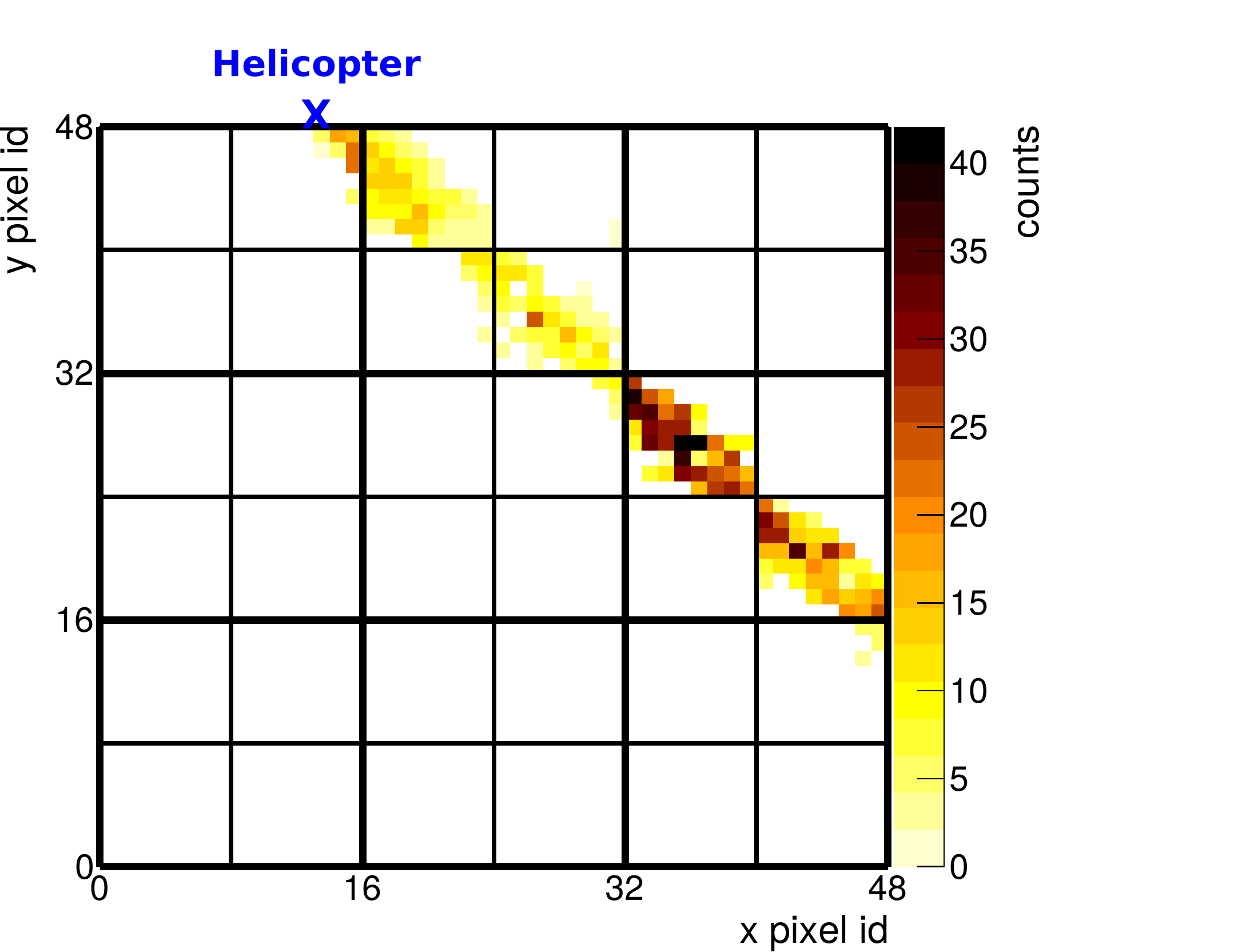}
   \caption{Laser track in the PDM at 05:29:25 UTC. Color represents the relative charge. The blue X is the helicopter position at this time.}
   \label{fig:ExampleTrack}
\end{figure}

The laser energy was switched every two minutes between $\sim$\SI{15}{\milli\J} and $\sim$\SI{10}{\milli\J}. The nominal laser energy corresponds to an EAS energy of about \SI{60}{\exa\eV} for \SI{15}{\milli\J}. A \SI{19}{\Hz} repetition rate was chosen to obtain a chance overlap at regular intervals between the \SI{20}{\Hz} readout of the balloon and the laser. This arrangement worked around a problem with the clock synchronization between the two systems, possibly caused by a faulty GPS antenna on the balloon.
The readout length is \SI{320}{\micro\second}. Assuming a minimum track length of the laser in the detector of 4 GTUs a chance overlap yields a number of potentially observable tracks of
\begin{equation*}
N = (0.32ms-0.01ms)\cdot 20Hz \cdot N_{\text{laser pulses}} = 0.0062 \cdot 105260 = 653
\end{equation*} 
This number has to be corrected for periods when we do not expect to identify any tracks due to cloud obscuration. In a first oder approximation the possibility for clear atmospheric conditions is 33\%. That gives an estimated number of 216 tracks. 
Out of these we were able to reconstruct 190. One reason for the lower number could be obstruction by clouds between the laser and the detector or positioning of the laser. The average energy of the shots, measured at the laser system, as a function of time can be seen in figure \ref{fig:LaserShots}. The energy is decreasing over time due to heating of the laser. The laser shots that were recorded by EUSO-balloon are superimposed. Most of the events were recorded when there were no clouds between the laser and the balloon.\\
An example of a recorded laser track can be found in figure \ref{fig:ExampleTrack}.

\section{Analysis of laser events}
\label{sec:Analysis}

\subsection{Track identification}
\label{subsec:IdTrack}
A track is a set of pixels from multiple GTUs. To identify tracks in the externally triggered data, a two-level pseudo-trigger was implemented in the offline data analysis. The first level is a simple threshold trigger. The average background per pixel over 128 GTUs is calculated. If at least 25 pixels have a value 5 sigmas above this average background the event is selected for further analysis. The event will be processed by the second level trigger which is looking for tracks. To determine if the set of pixels above background form a track two algorithms are available: one based on nearest neighbors and one on a linear time fit. Both algorithms use clusters of triggered pixels. A pixel is added to the cluster if its signal is 5 sigmas above its background and it is not further away than three pixels. The first algorithm adds a cluster to the track selection if it is next to another cluster in space and time. In case multiple tracks are found by this algorithm, the longest one is chosen. This method finds 205 tracks in the data. The second algorithm is more complex. It performs a linear time fit on the clusters to identify tracks in the data. 205 tracks are found with this algorithm as well. This latter algorithm is used for the work presented here.
\subsection{Algorithm for geometric reconstruction}
\label{subsec:Algo}
\begin{figure}
 \centering
 \includegraphics[width=.75\columnwidth]{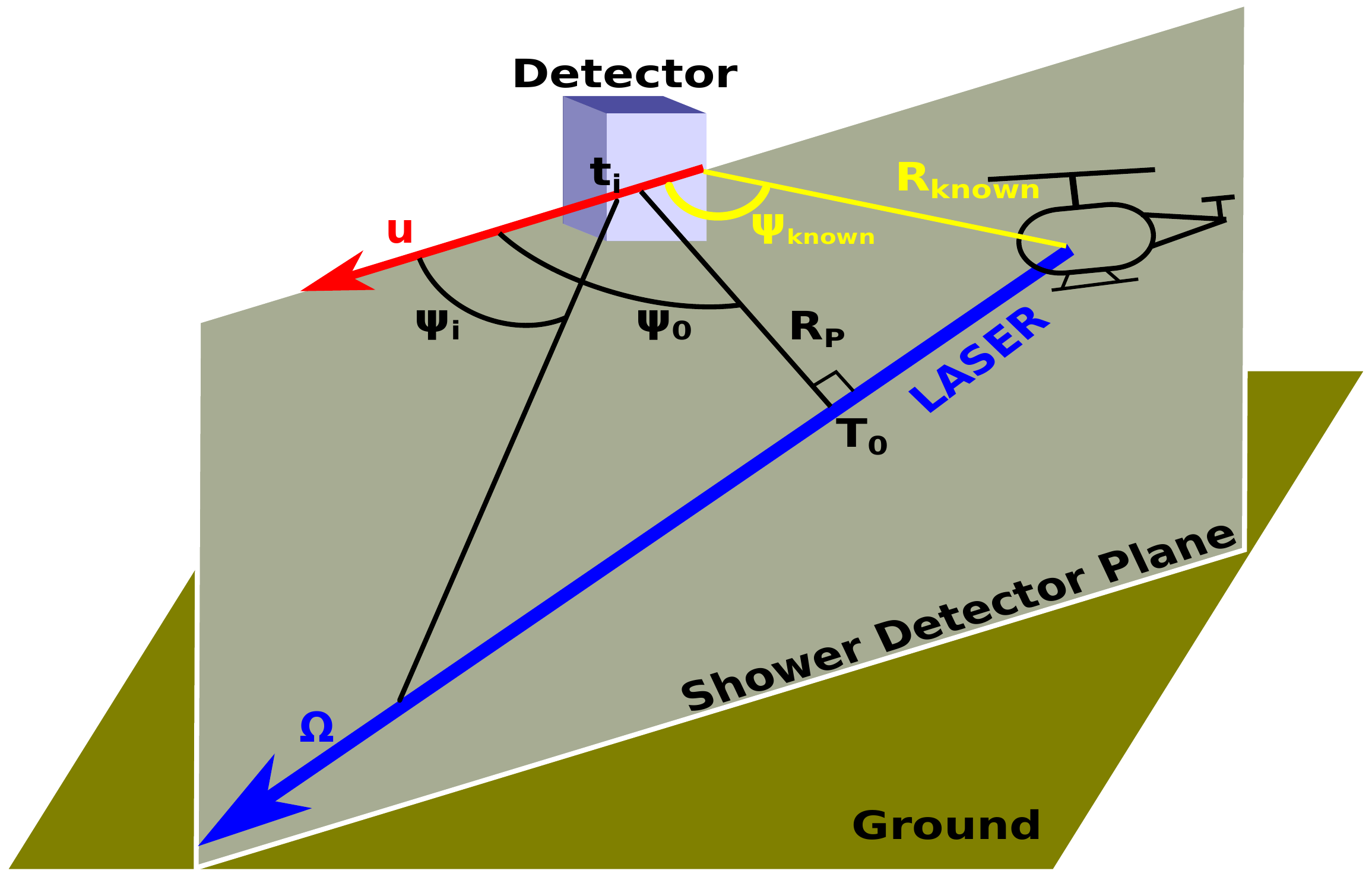}
 \caption{Illustration of the reconstruction of the geometrical direction of the laser tracks fired from the helicopter using the observables from the balloon. The parameters are explained in the text.}
 \label{fig:method}
\end{figure}

In this section, we explain the algorithm used to determine the geometry of the laser tracks.
The analysis follows two major steps. First, the pointing direction of the selected pixels is used to find the Shower Detector Plane (SDP) (Fig. \ref{fig:method}). In our case the shower is a laser track. The SDP is given by the location of the detector and the line of the shower axis. The norm vector of the SDP, $\hat{\textnormal{\textbf{SDP}}}$ is given by the pointing direction of the selected pixels and weighted by their count.
\begin{equation}
\hat{\textnormal{\textbf{SDP}}} = \dfrac{1}{\sqrt{(\sum_{i; j>i} C_i C_j \textbf{n}_i \times \textbf{n}_j)^2}}\sum_{i; j>i} C_i C_j \textbf{n}_i \times \textbf{n}_j,
\label{eq:sdp}
\end{equation}
where $C_i$, $n_i$ being, respectively, the charge and a unit vector along the pointing direction of the i$^{th}$ pixel in the track. $\textnormal{\textbf{u}}$, in the figure, is a unit vector lying in the SDP that is pointing in the horizontal direction. 

To reconstruct the direction of the event in the SDP, a trial nominal direction is estimated. Then the expected time for the signal to reach the detector is calculated for each pixel based on the region of the event axis to which it points (see Eq. \ref{eq:3parameter}). The difference between the expected and the observed time is compared and the parameters are adjusted to minimize time differences across the detector using the $\chi^2$ minimization method. The geometry with the minimum difference is used to reconstruct the shower axis. The distance of closest approach, $R_P$, and the angle from \textbf{u} to $R_P$, $\psi_0$ are the two parameters describing the axis. The arrival time at the i$^{\text{th}}$ pixel is given by

\begin{equation}
 \label{eq:3parameter}
  t_{i, expected} = T_0 + \dfrac{R_P}{c} \tan\left(\dfrac{\pi}{4}+\dfrac{\psi_0 - \psi_i}{2}\right)
\end{equation}
where $\psi_i=\textnormal{acos}(\textnormal{\textbf{u}}\cdot \textnormal{\textbf{n}}_i)$ is the pointing direction of each participating pixel in the SDP. $T_0$ is the time when the shower front reaches $R_P$. If the change in angular speed $d\psi/dt$ is small over the observed track length (low curvature in time vs. angle), the uncertainties in the 3 parameter fit can be large.\\

A constrained fit can be performed using a known position along the track, for example the source position. This reduces the number of parameters in Eq.\eqref{eq:3parameter} from 3 to 2 ($T_0$ and $\psi_0$)  given by

\begin{equation}
  \label{eq:2parameter}
  t_{i, ex} = T_0 + \dfrac{R_{\text{known}}}{c} \cdot \cos(\psi_{\text{known}} - \psi_0)\cdot \tan(\dfrac{\pi}{4} - \dfrac{\psi_i - \psi_0}{2})
\end{equation} 
where $R_{\text{known}}$ is the distance between the detector and the known source point (in our case, the helicopter position), and $\psi_{\text{known}}$ is the angle between the horizontal \textbf{u} and the known source point. This angle as well as $R_{\text{known}}$ are calculated based on the GPS positions of the detector and the helicopter.\\  
Finally, the laser reconstructed direction is given by
\begin{equation}
\bm{\Omega} = \sin(\psi_0)\hat{\textnormal{\textbf{u}}}+\cos(\psi_0) \left(\hat{\textnormal{\textbf{u}}} \times\hat{ \textnormal{\textbf{SDP}}}\right),
\label{eq:laser_direction} 
\end{equation}
where, the vector $\hat{\textnormal{\textbf{u}}}$ is contained in the SDP. It is therefore, perpendicular to the $\hat{\textnormal{\textbf{SDP}}}$, and can be simply taken as $\textbf{u}=(\hat{\textnormal{\textbf{SDP}}}_y,\;-\hat{\textnormal{\textbf{SDP}}}_x,\;0)$. The approach described here is also valid for the direction reconstruction of extensive air showers.

\subsection{Results of the direction reconstruction}
\label{subsec:Results}
\begin{figure}
  \begin{subfigure}{.53\columnwidth}
   \center
   \includegraphics[width=\columnwidth]{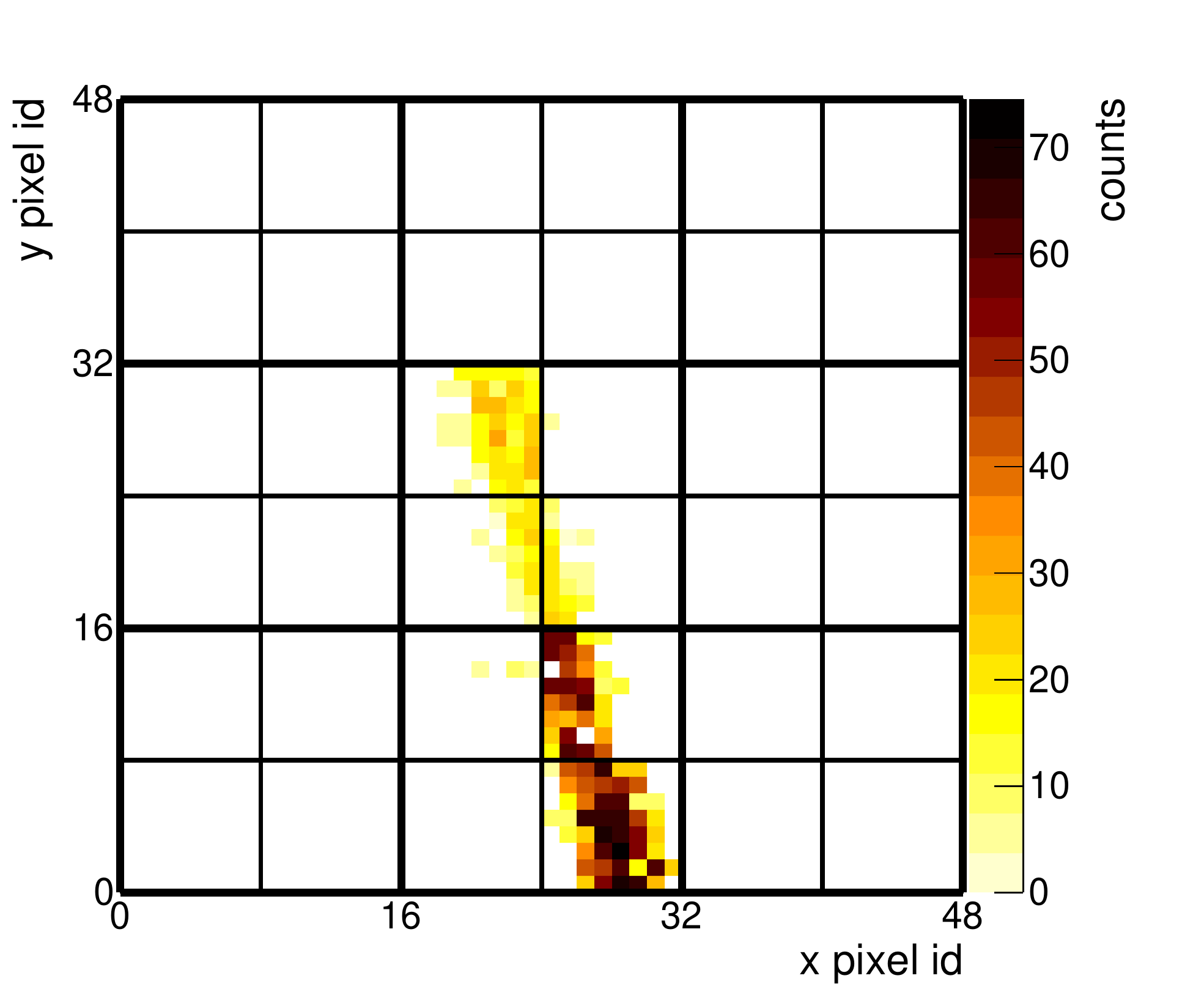}
   \caption{Laser track in the PDM. Color represents the relative charge.}
   \label{fig:ExampleTrack2}
   \end{subfigure}
   \hfill
   \begin{subfigure}{.45\columnwidth}
    \center
    \includegraphics[width=\columnwidth]{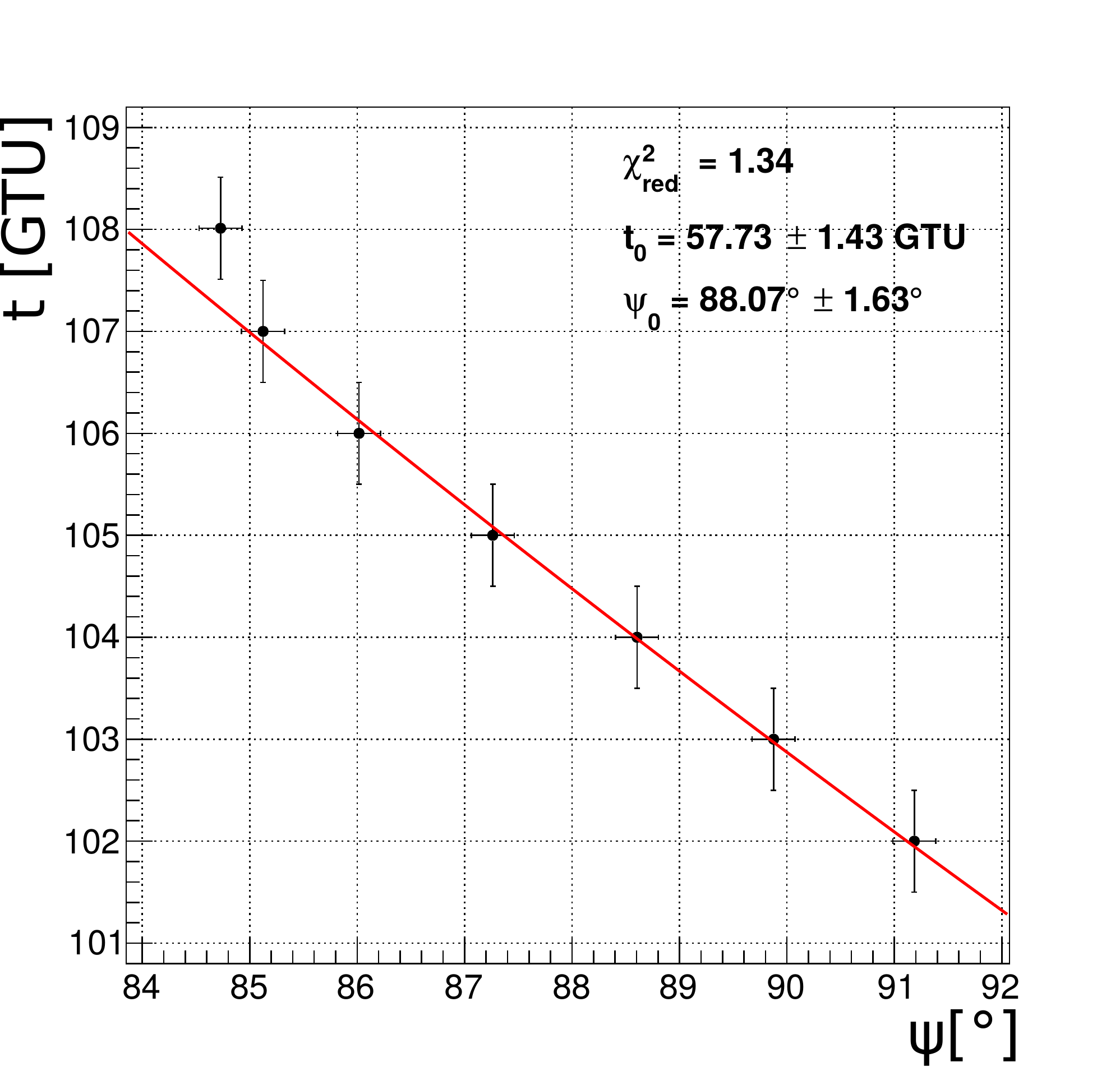}
    \caption{Timing profile of a laser event with constant 0.5 GTU timing error.}
    \label{fig:TimeFit}
   \end{subfigure}
   \caption{Example laser track at 05:40:24 UTC with corresponding time profile and fit.}
   \label{fig:exampleEvent}
\end{figure}

The laser tracks from the underflight have been analyzed to reconstruct the laser direction relative to the detector. An example track is shown in fig. \ref{fig:exampleEvent}, with the corresponding two-parameter timing fit (see equation \ref{eq:2parameter}). In this case, the fit was constrained using the position of the helicopter.\\

\begin{figure}
   \center
   \includegraphics[width=.75\columnwidth]{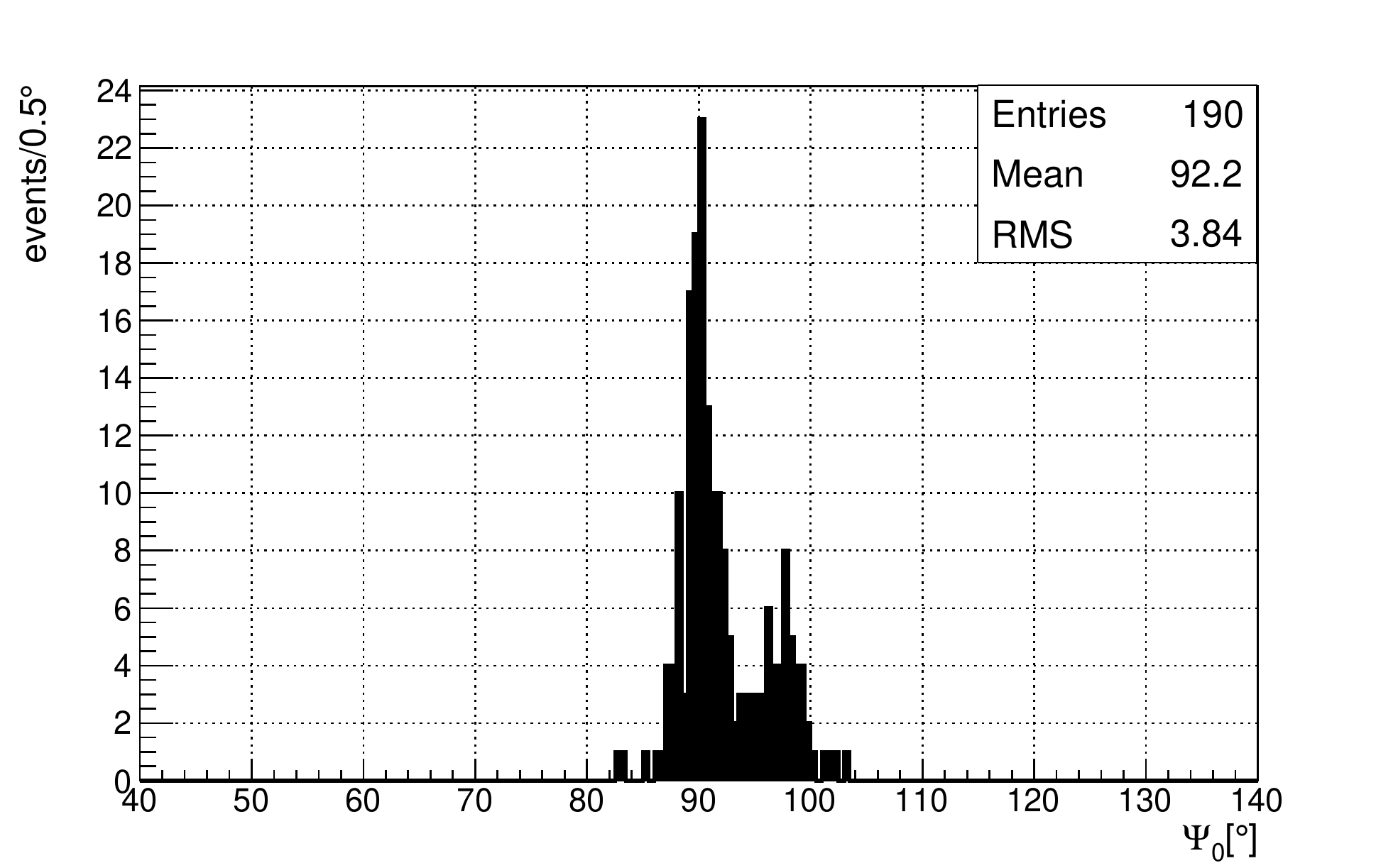}
   \caption{Zenith angle reconstruction of the helicopter laser shots with the 2-parameter fit method including only tracks with 4 GTUs or more.}
   \label{fig:2parameter}
\end{figure}

The instrument captured 205 laser track candidates. To ensure that the candidate is indeed a track and not a false positive of the track finding algorithm, we required a track length of at least 4 GTUs which corresponds to at least 2 degrees of freedom in our fit. Since the tracks were nominally horizontal, the track length is directly related to the observation duration expressed as the number of GTUs. This criterion reduces the number of tracks to 190. The reconstructed zenith angle is histogrammed in fig. \ref{fig:2parameter}.

\begin{figure}
   \begin{subfigure}{.48\columnwidth}
      \center
      \includegraphics[width=\columnwidth]{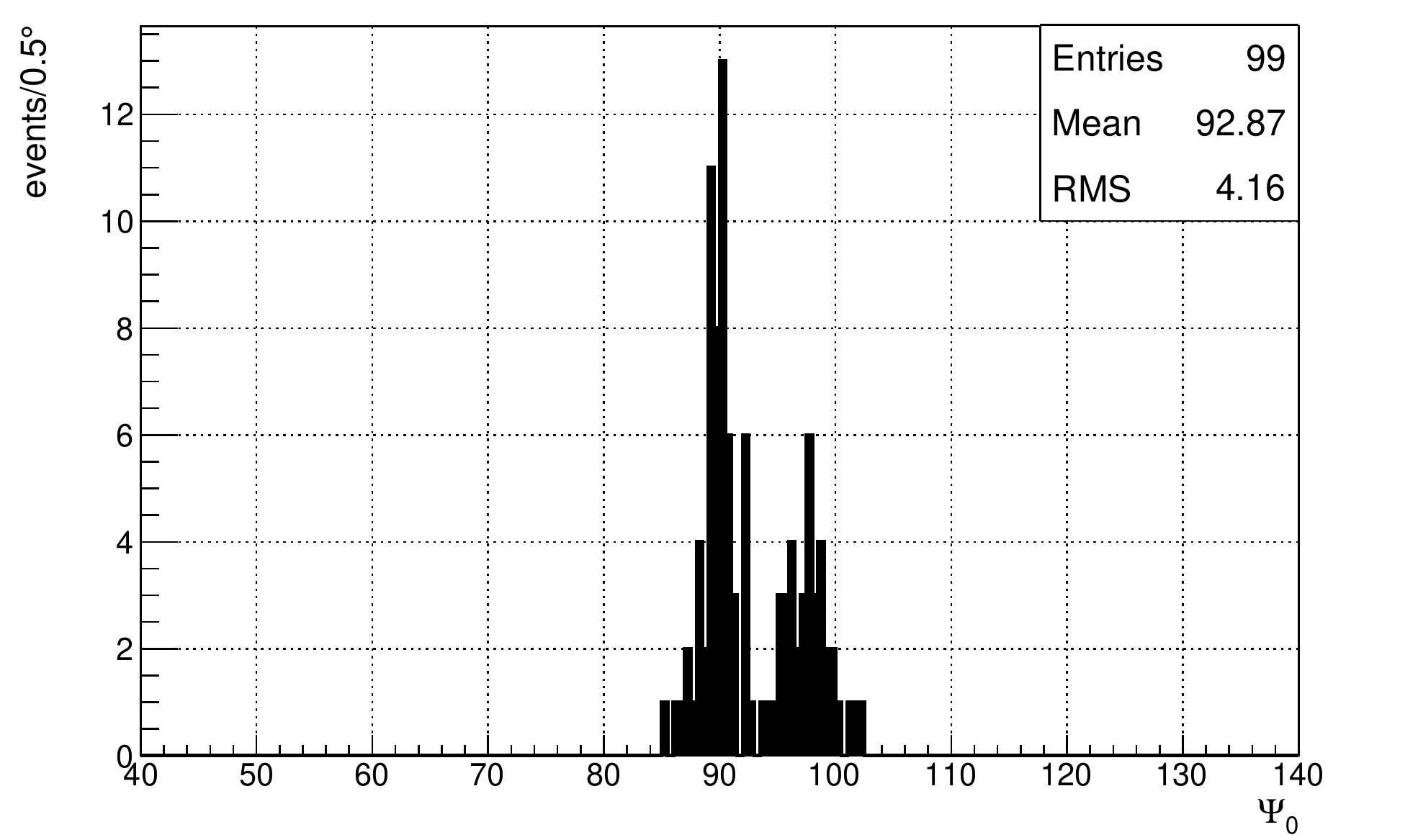}
      \caption{2-parameter reconstruction for an average laser energy of \SI{10}{\milli\J} (low setting).}
      \label{fig:lowE}
      \end{subfigure}
      \hfill
      \begin{subfigure}{.48\columnwidth}
       \center
       \includegraphics[width=\columnwidth]{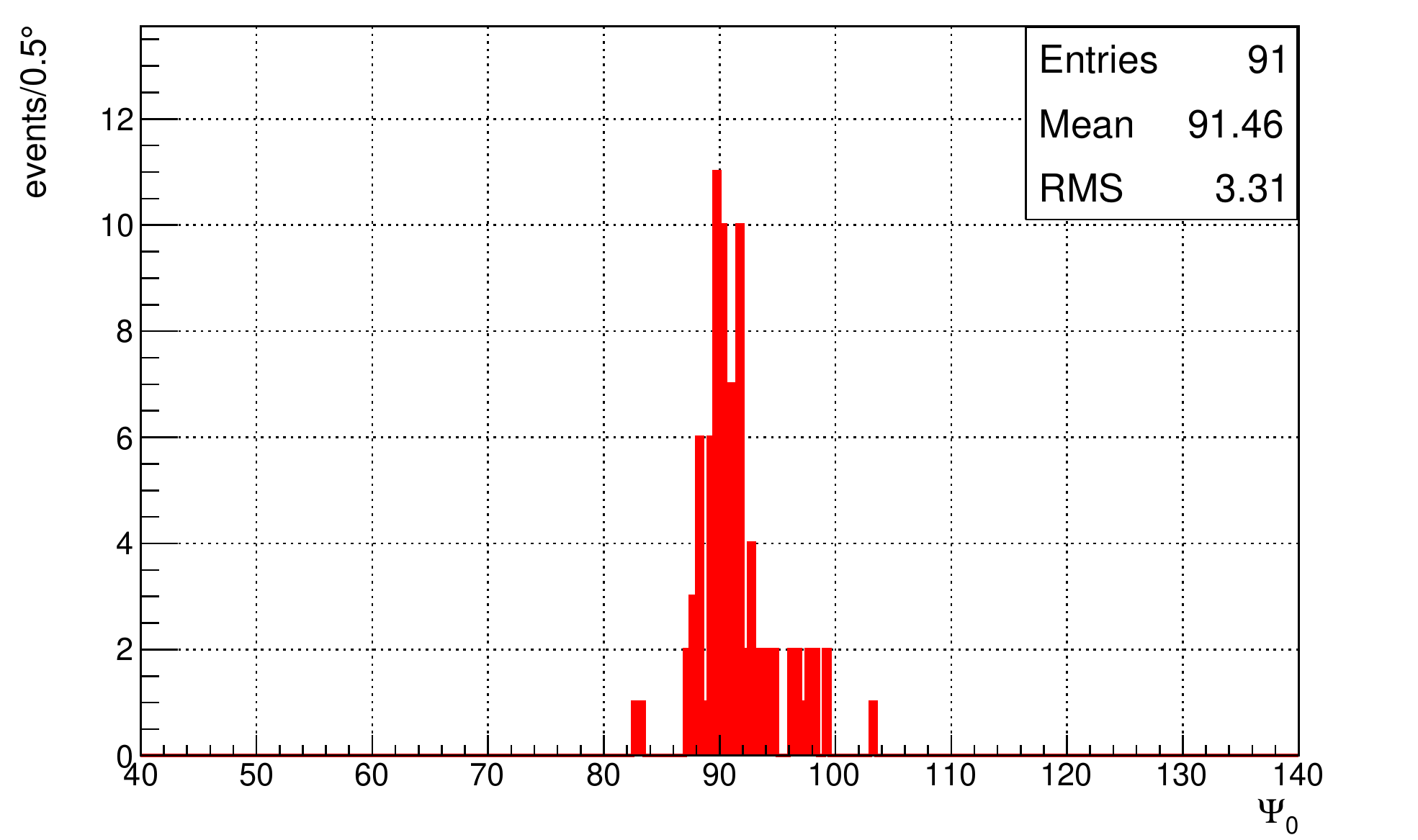}
       \caption{2 Parameter reconstruction for an average laser energy of \SI{15}{\milli\J} (high setting).}
       \label{fig:highE}
      \end{subfigure}
   \caption{Zenith angle reconstruction of the helicopter laser shots with the 2-parameter fit method split by laser energy. The minimum track duration is 4 GTUs.}
   \label{fig:low_high}
\end{figure}

The distribution of the 190 reconstructible events has a mean zenith angle of \unit[92.2]{$^\circ$} with a standard deviation of \unit[3.8]{$^\circ$}. The two populations visible are related to the two energy settings used for the laser. While the lower energy setting is contributing to both populations the higher setting only contributes to the population centered around \unit[90]{$^\circ$}. A possible explanation for this behavior is saturation of pixels inside the track, which shifts the weight of the timing fit. The result of splitting the dataset into high and low energy setting is shown in fig. \ref{fig:low_high}.\\
The expected mean value of the distribution should be slightly above \unit[90]{$^\circ$}. The reason is that the laser was mounted to produce horizontal tracks (meaning a zenith angle of \unit[90]{$^\circ$}) when the helicopter had a horizontal attitude. However, the helicopter was slightly turned sideways towards the ground by approximately \unit[1-2]{$^\circ$} to fly a circular pattern. The estimation of the bank angle uses the velocity of the helicopter and the assumption of truly circular flight pattern ($\theta = \arctan( v^2/(R\cdot g))$). The velocity and position of the helicopter were recorded using the on-board GPS at a 1 Hz rate. Using the position information it is possible to fit circles to the flight pattern and obtain an approximate radius for segments of the flight. 

\section{Conclusion}
\label{sec:conclusion}
The 2014 EUSO-Balloon flight made the first measurements of optical tracks by a fluorescence detector looking down on the atmosphere. The measurement required coordinating successfully the logistics of a balloon launch, a nighttime helicopter flight, on board laser and light sources, and GPS tracking. The laser tracks are simulating the signal of an EAS, proving the capability of EUSO-Balloon to observe such a signal.\\
Although the laser was too bright to perform an energy reconstruction (pixels were saturated), the beam direction was reconstructed relative to the detector. In the reconstruction it was assumed that the track is moving with the speed of light. The fact that the reconstructed zenith angles are reasonable insures that this assumption is true. The direction of the tracks could be reconstructed with a precision of \SI{\pm 3.8}{\degree}. The spread is less for the high energy setting of the laser than for the lower one. We note that the offset of \SI{\sim 2}{\degree} below horizontal is consistent with the estimated average tilt of the circling helicopter.\\
The obtained angular resolution is affected by various factors: EUSO-Balloon is a prototype instrument designed mainly to demonstrate the JEM-EUSO proof of principle. The detector was equipped with only two of the three Fresnel lenses in the original design for the optics, leading to a point spread function of around 9 pixels ($\backsimeq$ \unit[0.7]{$^\circ$}). In addition, the PDM had dead spots and only 30\% of the PDM was properly calibrated. Details of this calibration are presented in \cite{MorettoICRC2015}. The accuracy of the efficiency measurement of the MAPMTs with the highest gain (this mean around 30\% of the PDM) is better than 5\%. The \unit[2.5]{$\mu$s} resolution was too large for a reconstruction angular resolution within a few degrees at the short distance between the helicopter and the balloon of \unit[35]{km}.
This resolution does not represent the final resolution of the JEM-EUSO instrument. The distance between the detector and the shower will be around 10 times larger resolving the time resolution issue. The issues discovered during the mission led to an upgrade in the electronics used in the subsequent mission: EUSO-SPB1 \cite{ICRC2017:Lawrence}. Furthermore the experience will play an important role for the planned mission of EUSO-SPB2 \cite{SPB2}.
\section*{Acknowledgement}
The authors acknowledge strong support from the French Space Agency CNES who provided the leadership that made this achievement possible in a very short time as well as funding. We are indebted to the balloon division of CNES for a perfect launch, a smooth flight operation and flawless telemetry. The Canadian Space Agency has provided outstanding facilities at the Timmins Stratospheric Balloon Base and a quick and careful recovery of the instrument. We would like to thank NASA for their support in organizing and financing the helicopter underflight. 
This work was partially supported by the NASA grants NNX13AH55G, NNX13AH53G in the US. Japan is supported by the Basic Science Interdisciplinary Research Projects of RIKEN and JSPS KAKENHI Grant (22340063, 23340081, and 24244042), by the Italian Ministry of Foreign Affairs and International Cooperation, by the 'Helmholtz Alliance for Astroparticle Physics HAP' funded by the Initiative and Networking Fund of the Helmholtz Association, Germany, and by Slovak Academy of Sciences MVTS JEM-EUSO as well as VEGA grant agency project 2/0076/13. Russia is supported by the Russian Foundation for Basic Research Grant No 13-02-12175-ofi-m. The Spanish Consortium involved in the JEM-EUSO Space Mission is funded by MICINN \& MINECO under the Space Program projects: AYA2009-06037-E/AYA, AYA-ESP2010-19082, AYA-ESP2011-29489-C03, AYA-ESP2012-39115-C03,AYA-ESP2013-47816-C4, MINECO/FEDER-UNAH13-4E-2741, CSD2009-00064 (Consolider MULTIDARK) and by Comunidad de Madrid (CAM) under projects S2009/ESP-1496 \& S2013/ICE-2822.
We would like to thank the staff at our laboratories and home institutions for their strong and undivided support all along this project.

\noindent
{\Large\bf The JEM-EUSO Collaboration\\}
\vspace*{0.5cm}
\begin{sloppypar}
{\small \noindent 
G.~Abdellaoui$^{ah}$, 
S.~Abe$^{fq}$, 
J.H.~Adams Jr.$^{pd}$, 
A.~Ahriche$^{ae}$, 
D.~Allard$^{cc}$, 
L.~Allen$^{pb}$,
G.~Alonso$^{md}$, 
L.~Anchordoqui$^{pf}$,
A.~Anzalone$^{eh,ed}$, 
Y.~Arai$^{fs}$
% ...
}
\end{sloppypar}
\vspace*{.3cm}

{ \footnotesize
\noindent
% ...
$^{ae}$ Laboratory of Theoretical Physics LPT, University of Jijel, Jijel, Algeria \\
% ...
$^{ah}$ Telecom Lab., Faculty of Technology, University Abou Bekr Belkaid, Tlemcen, Algeria \\
% ...
$^{cc}$ APC, Univ Paris Diderot, CNRS/IN2P3, CEA/Irfu, Obs de Paris, Sorbonne Paris Cit\'e, France\\
% ...
$^{ed}$ Istituto Nazionale di Fisica Nucleare - Sezione di Catania, Italy\\
% ...
$^{eh}$ INAF - Istituto di Astrofisica Spaziale e Fisica Cosmica di Palermo, Italy\\
% ...
$^{fq}$ Nihon University Chiyoda, Tokyo, Japan\\ 
% ...
$^{fs}$ High Energy Accelerator Research Organization (KEK), Tsukuba, Japan\\ 
% ...
$^{md}$ Universidad Polit\'ecnia de madrid (UPM), Madrid, Spain\\
% ...
$^{pb}$ University of Chicago, USA\\
% ...
$^{pd}$ University of Alabama in Huntsville, Huntsville, USA\\
% ...
$^{pf}$ Lehman College, City University of New York (CUNY), USA\\
% ...
}


\end{document}